\documentstyle[12pt]{article}

\begin{document}

{ \pagestyle{empty}

\rightline{NWU-11/00}
\rightline{March 2000}
\rightline{~~~~~~~~~~}

\vskip 10mm

\centerline{\Large \bf The Scenario for the Astrophysics}
\centerline{\Large \bf with Scalar Field and the Cosmological Constant}

\vskip 20mm

\centerline{Masakatsu Kenmoku \footnote{E-mail address:
			kenmoku@phys.nara-wu.ac.jp}  and 
			Mayumi Ohto \footnote{E-mail address:
			oto@phys.nara-wu.ac.jp}}
\centerline{\it Department of Physics}
\centerline{\it Nara Women's University, Nara 630-8506, Japan}

\vskip 0.6cm

\centerline{Kazuyasu Shigemoto \footnote{E-mail address:
			shigemot@tezukayama-u.ac.jp} and
			Kunihiko Uehara\footnote{E-mail address:
			uehara@tezukayama-u.ac.jp}}
\centerline{\it Department of Physics}
\centerline{\it Tezukayama University, Nara 631-8501, Japan}

\vskip 1cm

\centerline{\bf Abstract}

In order to give the standard scenario of the astrophysics,
we study the Einstein theory with minimally 
coupled scalar field and the cosmological term by considering 
the scalar field as a candidate of the dark matter.
We obtained the exact solution in the cosmological scale and 
the approximate gravitational potential in the 
galactic or solar scale. 
We find that the scalar field plays the role of the dark matter
in some sense in the cosmological scale but it does not play
the role of the dark matter in the galactic or solar scale
within our approximation.

\vskip 4mm

\noindent
PACS number(s): 98.80.-k, 98.80.Hw, 97.60.Lf, 95.35.+d 
\hfil
\vfill

\newpage}

\renewcommand{\theequation}{\thesection.\arabic{equation}}
\setcounter{equation}{0}
\section{Introduction}
\indent

Recently the existence of the cosmological constant becomes
quite probable from the observation of the deep 
galaxy survey~\cite{Perlmutter,Riess}.
While the dark matter is necessary in various observations such as
the rotational curve of the spiral galaxy or the missing of
the ordinary matter in the cosmological scale.

Then we start from the theory of general relativity with
the dark matter and the cosmological constant in order to study
the standard scenario of the astrophysics, that is, the physics of
the cosmological, the galactic or solar scale.
Though the neutrino is the promising candidate of the dark matter, 
there is no established direct observation of the dark matter
as the ordinary matter.
There is another attempt to explain the rotation curves in
the theory of the Brans-Dicke theory~\cite{B-D,Fujii1}, where
the Newtonian force is modified by the effect of the scalar field.
In this paper, we consider the scalar field as a candidate 
of the dark matter~\cite{Guzman} together with the cosmological
constant and study their effects on time development of
the scale factor of the universe in the cosmological scale 
and the gravitational potential in the galactic or solar scale.

The famous scalar-tensor theory is the Brans-Dicke 
theory, but we adopt the Einstein theory with 
minimally coupled scalar field instead of 
the Brans-Dicke theory. Our principle of the choice of the theory
is the following. 
For the scalar-tensor gravity theory, we can transform one
from the Jordan frame to the Einstein frame by the conformal 
transformation~\cite{Jordan}. We prefer to adopt the Einstein frame 
because the post-Newtonian test of the general relativity 
such as the radar echo delay is quite stringent~\cite{Weinberg,Ours}. 
Also we prefer the Einstein theory with minimally coupled
scalar field because the scalar field as the dark matter means that
the scalar field has no direct interaction of the gravitational field
with the ordinary matter.


\setcounter{equation}{0}
\section{Classical Solution with Minimally Coupled Scalar and 
Cosmological Constant}
\subsection {Einstein Theory vs. Brans-Dicke Theory}
\indent

The Brans-Dicke theory~\cite{B-D,Fujii1} is the 
typical scalar-tensor theory of the gravity. The action of the Brans-Dicke 
theory with the cosmological term is given by 
\begin{eqnarray}
  I_{\rm BD}=\int d^4x \sqrt{-g} \left[
  \frac{1}{16\pi G}\left( \xi\phi^2R-2 \Lambda \phi^n \right)
  -\frac{1}{2}g^{\mu\nu}\partial_\mu\phi\partial_\nu\phi
  +{\cal L}_{\rm ordinary \ matter} \right] . 
\nonumber
\end{eqnarray}

\noindent
Uehara-Kim~\cite{Uehara} found the general solution for $n=2$ and matter 
dominant case, and Fujii~\cite{Fujii2} has found the special solution for 
general $n$. 

Putting $g_{\mu \nu}(x)=\Omega^{-2}(x) g_{* \mu \nu}(x)$ with
 $\Omega(x)=\sqrt{\xi}\phi(x)$, we obtain the following 
action~\cite{Jordan}
\begin{eqnarray}
  \hskip5mm
  I_{\rm BD}=\displaystyle{ \int } d^4x \sqrt{-g}
    \left[\displaystyle{ \frac{1}{16 \pi G}}
    \left( R_{*} -2 \Lambda e^{(n-4) \zeta\phi_{*}} \right)
   -\frac {1}{2} g_{*}^{ \mu \nu} \partial_{\mu}\phi_{*} 
    \partial_{\nu}\phi_{*} \right.
    \left. +\xi^{-2} e^{-4 \zeta\phi_{*}}{\cal L}_{*{\rm ordinary \ matter}}
   \right]~, 
\nonumber
\end{eqnarray}
where $\phi=\exp(\zeta\phi_{*})$ with
$\zeta^{-1}=\sqrt{1/\xi +3/4 \pi G}$
and ${\cal L}_{*{\rm ordinary \ matter}}$ is obtained from 
${\cal L}_{\rm ordinary \ matter}$ by replacing the metric part 
in the form $g_{\mu \nu} \rightarrow g_{\mu \nu}
\xi^{-1} \exp( \zeta\phi_{*}) $. 

Our philosophy to fix the theory comes from the following two principles: 
i) the kinetic part of the gravity is of the standard Einstein form,
because of the stringent constraint of the post-Newtonian test such as the 
delay of the radar echo experiment, 
ii) the scalar field has no direct coupling to the ordinary matter 
    nor gives the effect on the geodesic equation of the particle.
From these principles, we do not adopt the Brans-Dicke theory.  
In the following, we adopt the Einstein theory with standard cosmological 
term and the minimally coupled scalar field.

\subsection {Einstein Theory with Minimally Coupled Scalar Field}
\indent

By considering the minimally coupled scalar field 
as some kind of dark matter, we study the prototype of the 
time development of the scale factor of the universe in the cosmological  
scale and the gravitational potential in the galactic or solar scale.

We use Misner-Thorne-Wheeler notation~\cite{Wheeler} and consider 
Einstein action with the cosmological constant, the minimally coupled 
scalar field and the ordinary matter
\begin{eqnarray}
  I = \displaystyle{ \int } d^4x \sqrt{-g} \left[ 
   \displaystyle{ \frac{1}{16 \pi G}} \left(R -2 \Lambda \right)
  -\frac{1}{2} g^{\mu \nu} \partial_{\mu}\phi\partial_{\nu}\phi
   +{\cal L}_{\rm ordinary \ matter} \right]~, 
\label{e1}
\end{eqnarray}

\noindent
where $G$\ is the gravitational constant, $R$\ is the scalar curvature 
and $\phi$ is the minimally coupled scalar field. 
The equations of motion in this system are given by 
\begin{eqnarray}
  R_{\mu \nu}-{1 \over 2} g_{\mu \nu} R + \Lambda g_{\mu \nu}
  &=&8 \pi G (T^\phi_{\mu \nu}
            +T_{\mu \nu}) , 
\label{e2} \\
   \partial_{\mu}(\sqrt{-g} g^{\mu \nu} \partial_{\nu}\phi)&=&0,
\label{e3}
\end{eqnarray}

\noindent
where $T^\phi_{\mu \nu}
  =\partial_{\mu}\phi\partial_{\nu}\phi
    - \frac{1}{2} g_{\mu \nu} g^{\rho \sigma} 
      \partial_{\rho}\phi\partial_{\sigma}\phi$
and $T_{\mu \nu}$ is the energy-momentum tensor of the ordinary matter.


\setcounter{equation}{0}
\section{Cosmological Exact Solution}
\indent

In order to study classical solutions in cosmology,
we substitute the homogeneous, isotropic and flat metric
\begin{equation}
   ds^2=-{dt}^2+a(t)^2\left[ {dr}^2 
   +r^2 ({d \theta}^2 +\sin{\theta}^2 {d \varphi }^2) \right]~,  
\label{e4}
\end{equation}

\noindent
and  the perfect fluid expression of the ordinary matter 
$T_{\mu\nu}=(\rho+p)u_\mu u_\nu+p g_{\mu\nu}$
into equations of motion.
We denote $\rho $ and $p$ as the density and the pressure of the 
perfect fluid respectively and we can take $u_{\mu}=(1,0,0,0)$ in 
the co-moving system.  
Then equations of motion to be solved become
\begin{eqnarray}
    &&\left(\frac{\dot{a}}{a}\right)^2 
    -\frac{\Lambda}{3}=\frac{8 \pi G}{3} 
    \left( \rho +\frac{\dot{\phi}^2}{2} \right),
\label{e5} \\
     &&\left(\frac{\dot{a}}{a}\right)^2 
     +2 \frac{\ddot{a}}{a} - \Lambda
     =-8 \pi G  \left(p + \frac{\dot{\phi}^2}{2} \right),
\label{e6} \\
    && \frac{\ddot{\phi}}{\dot{\phi}} +3 \frac{\dot{a}}{a}= 0.
\label{e7}
\end{eqnarray}

\noindent
We consider the perfect fluid characterized by 
$ p =\gamma \rho $, and we obtain the conservation law of the 
ordinary matter density by taking the linear combination 
of Eqs.(\ref{e5}), (\ref{e6}) and (\ref{e7}). From Eq.(\ref{e7}), 
we have another conservation law. Then we have the following two 
conservation laws 
\begin{eqnarray}
  && \rho = \rho_{0} a^{-3(1+\gamma)}, 
\label{e8}\\
  && \dot{\phi}=\frac{k}{a^3}, 
\label{e9}
\end{eqnarray}

\noindent
where $\rho_0$ and $k$ are integration constants.
The equation to be solved becomes Eq.(\ref{e5}) with the 
conditions Eqs.({\ref{e8}) and (\ref{e9}).
Substituting Eqs.({\ref{e8}) and (\ref{e9}) into
Eq.({\ref{e5}), we have the equation of the form 
\begin{eqnarray}
    && \left(\frac{\dot{a}}{a}\right)^2 
    -\frac{\Lambda}{3}=\frac{8 \pi G}{3} 
    \left( \frac{\rho_{0}}{a^{3(1+\gamma)}}+\frac{k^2}{2a^6} \right).
\label{e10} 
\end{eqnarray}

As the mathematical problem, we can solve exactly in the $\gamma=1$ 
and $\gamma=0$ case, but $\gamma=1$ case is unphysical.
Then we consider only the $\gamma=0$ case, that is, the matter dominant case.
In this case, 
we obtain 
\begin{eqnarray}
    && \left(\frac{\dot{a}}{a}\right)^2 
    -\frac{4 \pi G}{3} \left( \dot{\phi} +\frac{\rho_0}{k} \right)^2
     =\frac{\Lambda}{3} -\frac{4 \pi G {\rho_0}^2 }{3 k^2}
\label{e11} 
\end{eqnarray}

\noindent
by using Eqs.(\ref{e5}), (\ref{e8}) and (\ref{e9}).

\vskip 2mm
\subsection{$k^2 \Lambda > 4 \pi G {\rho_0}^2$ case  }
\indent

In this case the cosmological and/or the scalar term are dominant,
and we parametrize 
\begin{eqnarray}
    && \frac{\dot{a}}{a}
    =\sqrt{ \frac{\Lambda}{3} -\frac{4 \pi G {\rho_0}^2 }{3 k^2}} 
    \cosh {\Theta},
\label{e12}\\
    && \dot{\phi}
    =    -\frac{\rho_0}{k}
  +\sqrt{ \frac{\Lambda}{ 4 \pi G }-\frac{{\rho_0}^2}{k^2}} 
\sinh {\Theta}.
\label{e13}
\end{eqnarray}

\noindent
Substituting this parametrization into Eq.(\ref{e7}), we have 
\begin{eqnarray}
  && \dot{\Theta}+\sqrt{\frac{3 \Lambda}{1+A^2}}
  (\sinh{\Theta}-A)=0, 
\label{e14}
\end{eqnarray}

\noindent
where $A^{-1}=\sqrt{k^2 \Lambda/4 \pi G {\rho_0}^2-1}$.
Then we obtain 
\begin{eqnarray}
 && \frac{1}{\sqrt{1+A^2}} \log 
  \left| \frac{-A \tanh (\Theta/2)-1+\sqrt{1+A^2}}
 {-A \tanh (\Theta/2)-1-\sqrt{1+A^2}} \right|
  =-\sqrt{\frac{3 \Lambda}{1+A^2}} (t-t_0)
\label{e15}
\end{eqnarray}

\noindent
by using the formula 
\begin{eqnarray}
\int \frac{d \Theta}{\sinh{\Theta}-A}
=\frac{1}{\sqrt{1+A^2}} 
\log \left| \frac{-A \tanh (\Theta/2)-1+\sqrt{1+A^2}}
   {-A \tanh (\Theta/2)-1-\sqrt{1+A^2}} \right|.
\label{e16}
\end{eqnarray}

\noindent
And then we have the relation 
\begin{eqnarray}
 && \frac{-A \tanh (\Theta/2)-1+\sqrt{1+A^2}}
    {-A \tanh (\Theta/2) -1-\sqrt{1+A^2}}
     =\exp \left(- \sqrt{3 \Lambda} \left(t-t_0\right)\right),
\label{e17}
\end{eqnarray}

\noindent
which gives the relation
\begin{eqnarray}
 && \tanh (\Theta/2)=-\frac{1}{A}+\frac{\sqrt{1+A^2}}
    {A \tanh \left(\sqrt{3 \Lambda}\left(t-t_0\right)/2 \right)}.
\label{e18}
\end{eqnarray}

\noindent
Using this relation, we can write $\cosh{\Theta}$ and
$\sinh{\Theta}$ in the form 
\begin{eqnarray}
  &&\hskip-8mm
  \cosh{\Theta}
=\frac{1+\tanh^2 (\Theta/2)}{1-\tanh^2 (\Theta/2)}
  =\frac{(1+A^2) \cosh \left( T-T_0 \right)- \sqrt{1+A^2} 
   \sinh \left( T-T_0 \right) }
   {-A^2-\cosh \left( T-T_0 \right) +\sqrt{1+A^2} 
   \sinh \left(T-T_0 \right)},
\label{e19}\\
  &&\hskip-8mm
  \sinh{\Theta}
=\frac{2 \tanh (\Theta/2)}{1-\tanh^2 (\Theta/2)}
  =\frac{A \left( 1-\cosh \left(T-T_0 \right)
   + \sqrt{1+A^2} \sinh \left(T-T_0 \right)  \right)}
   {-A^2-\cosh \left( T-T_0 \right) +\sqrt{1+A^2} 
    \sinh \left( T-T_0 \right) },
\label{e20}
\end{eqnarray}

\noindent
where $T = \sqrt{3 \Lambda} t$ and $T_0=\sqrt{3 \Lambda} t_0$.
Introducing $\Theta_0$ through the relation 
$\cosh{\Theta_0}=\sqrt{1+A^2}/A$, $\sinh{\Theta_0}=1/A$, we 
can simplify the above expression in the form 
\begin{eqnarray}
  && \cosh{\Theta}=\frac{\sqrt{1+A^2} \cosh (T-T_0 -\Theta_0 )}
                  {\sinh (T-T_0 -\Theta_0 )-A},
\label{e21}\\
  && \sinh{\Theta}=A \left(1+\frac{(A+1/A)}
                 {\sinh (T-T_0 -\Theta_0 )-A} \right).
\label{e22}
\end{eqnarray}

Using Eqs.(\ref{e12}) and (\ref{e21}), we have 
\begin{eqnarray}
  \log{a}&=&\int \frac{d a}{a}
  =\sqrt{ \frac{\Lambda}{3}-\frac{4 \pi G {\rho_0}^2}{3 k^2}}
    \int dt \cosh{\Theta} \nonumber \\
  &=&\frac{1}{3}\sqrt{1-\frac{4 \pi G {\rho_0}^2}{k^2 {\Lambda}}}
   \sqrt{1+A^2}  \int d T \frac{\cosh \left(T-T_0 -\Theta_0 \right)}
   {\sinh \left( T-T_0 -\Theta_0 \right)-A }  \nonumber \\
  &=&\frac{1}{3} \log \left| \sinh ( T-T_0 -\Theta_0 ) - A 
    \right| + {\rm const.} ,
    \label{e23}
\end{eqnarray}

\noindent
where we use the relation $4\pi G{\rho_0}^2/k^2 \Lambda=A^2/(1+A^2)$.
Therefore we have
\begin{eqnarray}
  a(t)=a_0 \Big( \sinh ( T-T_0 -\Theta_0 )-A \Big)^{1/3},
\label{e24}
\end{eqnarray}

\noindent
where $a_0$ is the constant.

Similarly, from Eqs.({\ref{e13}) and (\ref{e22}), we have 
\begin{eqnarray}
  \phi&=& \int dt \left(-\frac{\rho_0}{k}
   +\sqrt{ \frac{\Lambda}{4 \pi G}
-\frac{{\rho_0}^2}{k^2}} \sinh{\Theta}
    \right)
  =- \frac{ \rho_0 }{k \sqrt{3 \Lambda} } 
   \int dT \left(1-\frac{\sinh{\Theta}}{A} \right)
   \nonumber \\
  &=& \frac{\rho_0 (1+A^2)}{ k A \sqrt{3 \Lambda}} 
    \int \frac{dT} {\sinh (T-T_0 -\Theta_0 )-A } \nonumber \\
  &=&\phi_0
   + \frac{1}{\sqrt{12\pi G}} 
   \log \left| \frac{A \tanh \Big( (T-T_0 -\Theta_0 )/2 \Big)
                      +1-\sqrt{1+A^2}}
      {A \tanh \Big((T-T_0 -\Theta_0 )/2 \Big)+1+\sqrt{1+A^2}}
      \right|   \nonumber \\
  &=&\phi_1
   + \frac{1}{\sqrt{12\pi G}} 
   \log \left| \frac{\exp(T-T_0 -\Theta_0)-A- \sqrt{1+A^2}}
      {\exp(T-T_0 -\Theta_0 )-A + \sqrt{1+A^2}}\right| ,
\label{e25}
\end{eqnarray}

\noindent
where $\phi_0$ is the constant and $\phi_1$ is given by 
$\displaystyle{\phi_1=\phi_0+\frac{1}{\sqrt{12 \pi G}} 
\log \left| \frac{1+A-\sqrt{1+A^2}}{1+A+\sqrt{1+A^2}}\right|}$.

The integration constant $a_0$ is not the independent 
integration constant but it can be expressed by $k$ and $\rho_0$.
From Eqs.(\ref{e9}), (\ref{e24}) and (\ref{e25}), we have 
\begin{eqnarray}
  \dot{\phi}&=&\frac{\rho_0 (1+A^2)} 
  {k A \Big( \sinh (T-T_0-\Theta_0 )-A \Big) }  
\nonumber \\
  &=& \frac{\rho_0 (1+A^2) {a_0}^3 }{k A a^3}
   =\frac{k}{a^3}.
\label{e26}
\end{eqnarray}

\noindent
which gives the relation 
$a_{0}^3=k^2 A/\rho_0 (1+A^2)$. 
This can be written in the form 
\begin{eqnarray}
   k^2 \Lambda -4 \pi G {\rho_0}^2
=\frac{\Lambda^2 {a_0}^6}{4 \pi G}.
\label{e27}
\end{eqnarray}

\noindent
Then we can obtain 
\begin{eqnarray}
 a_0=\Big( 4 \pi G (\frac{k^2}{\Lambda}
	-\frac{4\pi G \rho_0^2}{\Lambda^2} ) \Big)^{1/6}.
\label{e28}
\end{eqnarray}

Therefore we have the exact solution in the form 
\begin{eqnarray}
  a(t)&=& \Big( 4 \pi G (\frac{k^2}{\Lambda}
	-\frac{4\pi G \rho_0^2}{\Lambda^2} ) \Big)^{1/6}
		\Big( \sinh(T-T_1 )-A \Big)^{1/3},
\label{e29}\\
  \phi(t)&=&\phi_1+\frac{1}{\sqrt{12 \pi G}}
   \log \left| \frac { \exp{(T-T_1)}-A-\sqrt{1+A^2}}
                     { \exp{(T-T_1)}-A + \sqrt{1+A^2}}  \right|,
\label{e30}
\end{eqnarray}

\noindent
where
\begin{eqnarray}
 T = \sqrt{3 \Lambda}t , \quad 
 T_1=T_0-\Theta_0=\sqrt{3 \Lambda} t_1={\rm const.},
 \nonumber \\
 A^{-1}=\sqrt{\frac{k^2 \Lambda}{4 \pi G \rho_{0}^2}-1},
 \quad \phi_1={\rm const.}. \nonumber
\end{eqnarray}

\vskip 2mm

\subsection{ $4 \pi G {\rho_0}^2 > k^2 \Lambda $ case  }
\indent

In this case the ordinary matter is dominant, and we 
parametrize 
\begin{eqnarray}
    && \frac{\dot{a}}{a}
    =\sqrt{ \frac{4 \pi G {\rho_0}^2 }{3 k^2}-\frac{\Lambda}{3} } 
    \sinh{\Theta},
\label{e31}\\
    && \dot{\phi}
    =-\frac{\rho_0}{k}
   +\sqrt{ \frac{{\rho_0}^2}{k^2}
-\frac{\Lambda}{4 \pi G}} \cosh{\Theta}.
\label{e32}
\end{eqnarray}

\noindent
Substituting this parametrization into Eq.(\ref{e7}), we have 
\begin{eqnarray}
  && \dot{\Theta}
+\sqrt{\frac{3 \Lambda}{B^2-1}}(\cosh{\Theta}-B)=0, 
\label{e33}
\end{eqnarray}

\noindent
where $B^{-1} = \sqrt{1-k^2 \Lambda/4 \pi G{\rho_0}^2}$.
Then we obtain 
\begin{eqnarray}
 && \frac{1}{\sqrt{B^2-1}} \log 
  \left| \frac{1-B+\sqrt{B^2-1}\tanh (\Theta/2)}
              {1-B-\sqrt{B^2-1}\tanh (\Theta/2)} \right|
  =- \sqrt{\frac{3 \Lambda}{B^2-1}} (t-t_0),
\label{e34}
\end{eqnarray}

\noindent
by using the formula 
\begin{eqnarray}
\int \frac{d \Theta}{\cosh{\Theta}-B}
=\frac{1}{\sqrt{B^2-1}} \log \left| 
  \frac{1-B+\sqrt{B^2-1} \tanh (\Theta/2)}
       {1-B-\sqrt{B^2-1} \tanh (\Theta/2)} \right|.
  \nonumber
\end{eqnarray}

\noindent
Then we have the relation 
\begin{eqnarray}
 && \frac{1-B+\sqrt{B^2-1} \tanh (\Theta/2)}
         {1-B-\sqrt{B^2-1} \tanh (\Theta/2)}
     = - \exp \left(- \sqrt{3 \Lambda} (t-t_0) \right),
\label{e35}
\end{eqnarray}

\noindent
where we take the branch of the logarithm in such a way as the scale 
factor of the universe behaves as the power law in time at the very early 
age of the universe. 
Then we have the relation
\begin{eqnarray}
 && \tanh (\Theta/2)=\sqrt{\frac{B-1}{B+1}} 
   \tanh \left(\sqrt{3 \Lambda}(t-t_0)/2 \right).
\label{e36}
\end{eqnarray}

\noindent
Using this relation, we can write $\sinh{\Theta}$ and 
$\cosh{\Theta}$ in the form 
\begin{eqnarray}
  \sinh{\Theta}&=&\frac{2 \tanh (\Theta/2)}
    {1-\tanh^2 (\Theta/2)}
  =\frac{ \sqrt{B^2-1} \sinh \left(T-T_0 \right)}
   {\cosh \left( T-T_0 \right)-B},
\label{e37}\\
  \cosh{\Theta}
&=&\frac{1+\tanh^2 (\Theta/2)}{1-\tanh^2 (\Theta/2)}
  =\frac{B \cosh \left( T-T_0 \right)-1}
   {\cosh \left( T-T_0 \right)-B },
\label{e38}
\end{eqnarray}

\noindent
where $T = \sqrt{3 \Lambda} t$ and $T_0=\sqrt{3 \Lambda} t_0$.

Using Eqs.(\ref{e31}) and (\ref{e37}), we have 
\begin{eqnarray}
  \log{a}&=&\int \frac{d a}{a}
  =\sqrt{ \frac{4 \pi G {\rho_0}^2}{3 k^2}-\frac{\Lambda}{3}}
    \int dt \sinh{\Theta} \nonumber \\
  &=&\frac{1}{3}\sqrt{\frac{4 \pi G {\rho_0}^2}{k^2 \Lambda}-1}
   \int d T \frac{ \sqrt{B^2-1} \sinh \left( T-T_0 \right) }  
   {\cosh \left( T-T_0 \right)-B}   \nonumber \\
  &=&\frac{1}{3} \log \left| \cosh \left( T-T_0 \right)-B \right|
      +{\rm const.},
\label{e39}
\end{eqnarray}

\noindent
where we use the relation 
$ 4 \pi G {\rho_0}^2/k^2 \Lambda=B^2/(B^2-1)$.
Therefore we have
\begin{eqnarray}
  a(t)=a_0 \Big( \cosh( T-T_0 ) -B \Big)^{1/3},
\label{e40}
\end{eqnarray}

\noindent
where $a_0$ is the constant and $(T-T_0)=\sqrt{ 3 \Lambda} (t-t_0)$.

Similarly, from Eqs.({\ref{e32}) and (\ref{e38}), we have
\begin{eqnarray}
  \phi &=& \int dt \left(
 \sqrt{1- \frac{k^2 \Lambda}{4 \pi G {\rho_0}^2 }} \cosh{\Theta}-1
    \right)
  =\frac{ \rho_0 }{k \sqrt{3 \Lambda} } 
   \int dT \left(\frac{\cosh{\Theta}}{B}-1 \right)
   \nonumber \\
  &=& \frac{\rho_0 (B^2-1)}{ k B \sqrt{3 \Lambda}} 
    \int \frac{dT} { \cosh \left(T-T_0 \right)-B } \nonumber \\
  &=&\phi_0
   + \frac{\rho_0 \sqrt{B^2-1} }{ k B \sqrt{3 \Lambda}} 
   \log \left| 
   \frac{1-B+\sqrt{B^2-1} \tanh \Big( ( T-T_0 )/2 \Big)}
        {1-B-\sqrt{B^2-1} \tanh \Big( ( T-T_0 )/2 \Big)}
                 \right|    \nonumber \\
  &=&\phi_1
   + \frac{1}{\sqrt{12 \pi G}} 
   \log \left| \frac{ \exp \left(T-T_0 \right) -B - \sqrt{B^2-1}}
               {\exp \left(T-T_0 \right) -B + \sqrt{B^2-1}}
                 \right|,
\label{e41}
\end{eqnarray}

\noindent
where $\phi_0$ is the constant and $\phi_1$ is given by
$\displaystyle{\phi_1=\phi_0+\frac{1}{\sqrt{12 \pi G}} 
  \log \left| \frac{1+B-\sqrt{B^2-1}}{1+B+\sqrt{B^2-1}}\right|}$.

The integration constant $a_0$ is not the independent 
integration constant but it can be expressed by $k$ and $\rho_0$.
From Eqs.(\ref{e9}), (\ref{e40}) and (\ref{e41}), we have 
\begin{eqnarray}
  && \dot{\phi}=\frac{(B^2-1) \rho_0} 
  {k B \Big( \cosh( T-T_0 ) -B \Big) }  \nonumber \\
  &&  =\frac{(B^2-1) \rho_0 {a_0}^3} {k B a^3} =\frac{k}{a^3},
\label{e42}
\end{eqnarray}

\noindent
which gives the relation 
$k^2=(B^2-1) \rho_0 {a_0}^3 / B$. This can be written in the form 
\begin{eqnarray}
   4 \pi G {\rho_0}^2-k^2 \Lambda 
=\frac{\Lambda^2 {a_0}^6}{4 \pi G}.
\label{e43}
\end{eqnarray}

\noindent
Then we can obtain 
\begin{eqnarray}
 a_0=\Big( 4 \pi G ( \frac{4 \pi G {\rho_0}^2}{\Lambda^2}
    -\frac{k^2}{\Lambda} ) \Big)^{1/6}.
\label{e44}
\end{eqnarray}

Therefore we have the exact solution in the form 
\begin{eqnarray}
  a(t)&=& \Big( 4 \pi G ( \frac{4 \pi G {\rho_0}^2}{\Lambda^2}
    -\frac{k^2}{\Lambda} ) \Big)^{1/6}
    \Big( \cosh( T-T_0 )-B \Big)^{1/3},
\label{e45}\\
  \phi(t)&=&\phi_0
   + \frac{1}{\sqrt{12 \pi G}} 
   \log \left| \frac{ \exp \left(T-T_0 \right) -B - \sqrt{B^2-1}}
               {\exp \left(T-T_0 \right) -B + \sqrt{B^2-1}}
                 \right|,
\label{e46}
\end{eqnarray}

\noindent
where
\begin{eqnarray}
   T = \sqrt{3 \Lambda} t, \quad  
   T_0=\sqrt{3 \Lambda} t_0={\rm const.},
   \nonumber \\
   B^{-1}= \sqrt{1-\frac{k^2 \Lambda}{4 \pi G {\rho_0}^2}}, \quad 
   \phi_0={\rm const.}\ . \nonumber
\end{eqnarray}

\vskip 2mm

\subsection{ $4 \pi G {\rho_0}^2 = k^2 \Lambda $ case  }
\indent

For the completeness of the solution, we give the exact solution in this case.
As the method to solve the equation is similar, we give only the result. 
The solution is given by 
\begin{eqnarray}
  a(t)&=&\left( \frac{4 \pi G \rho_0 }{\Lambda} \right)^{1/3}
    \Big( \exp(T-T_0 )-1 \Big)^{1/3},
\label{e47}\\
  \phi(t)&=&\phi_0
   + \frac{1}{\sqrt{12 \pi G}} 
   \log \left| 1-\exp \Big(-(T-T_0) \Big) \right|,
\label{e48}
\end{eqnarray}

\noindent
where
\begin{eqnarray}
   T = \sqrt{3  \Lambda} t, 
\quad T_0=\sqrt{3 \Lambda} t_0={\rm const.},
   \quad  \phi_0={\rm const.}\ . \nonumber
\end{eqnarray}

\vskip 2mm

\subsection{ Special Limiting Case}
\underline{i) $\rho_0=0$ case (no ordinary matter)}
\indent

In case there is no ordinary matter $\rho_0 \rightarrow 0$, 
which corresponds to  
$A \rightarrow \sqrt{4 \pi G {\rho_0}^2/ k^2 \Lambda}$, we 
have the expression 
\begin{eqnarray}
  a(t)&=& \left(\frac{4 \pi G k^2} {\Lambda} \right)^{1/6}
   \Big( \sinh(T-T_1 ) \Big)^{1/3},
\label{e49}\\
  \phi(t)&=&\phi_1+\frac{1}{\sqrt{12 \pi G}} 
   \log \left| \tanh \Big( (T-T_1 )/2 \Big) \right|. 
\label{e50}
\end{eqnarray}

\vskip 3mm
\noindent
\underline{ii) $k=0$ case (no scalar matter): Lema\^{i}tre 
universe~\cite{Lemaitre}}

In case there is no scalar matter $k \rightarrow 0$, 
which corresponds to  
$B \rightarrow 1$, we 
have the expression 
\begin{eqnarray}
  a(t)&=&\left(\frac{ 4 \pi G \rho_0 } {\Lambda} \right)^{1/3} 
     \Big( \cosh( T-T_1 )-1 \Big)^{1/3},
\label{e51}\\
  \phi(t)&=&\phi_1.
\label{e52}
\end{eqnarray}

\vskip 3mm
\noindent
\underline{iii) $\Lambda=0$ case (no cosmological constant)}

In case there is no cosmological term $\Lambda \rightarrow 0$, 
which corresponds to  
$B \rightarrow 1+k^2 \Lambda/8 \pi G {\rho_0}^2$,
we have the expression 
\begin{eqnarray}
  a(t)&=& \lim_{\Lambda \rightarrow 0}
     \left( \frac{4\pi G\rho_0}{\Lambda} \right)^{1/3} 
     \left( \frac{3 \Lambda \left(t-t_0 \right)^2 }{2}
       -\frac{k^2 \Lambda}{8 \pi G {\rho_0}^2}\right)^{1/3}
   \nonumber\\
  &=&\left( 6 \pi G \rho_0 \right)^{1/3}     
     \left( \left(t-t_0 \right)^2-\frac{k^2}{12 \pi G {\rho_0}^2} 
       \right)^{1/3}
\label{e53}\\
  \phi(t)&=&\lim_{ \Lambda \rightarrow 0} \left\{\phi_0
      + \frac{1}{\sqrt{12 \pi G}} \log \left|
   \frac{ \sqrt{3 \Lambda}(t-t_0)-k \sqrt{\Lambda}
        /\sqrt{4 \pi G \rho_0^2}}
    { \sqrt{3 \Lambda}(t-t_0)+k \sqrt{\Lambda}
       /\sqrt{4 \pi G \rho_0^2}}
      \right|  \right\}   \nonumber \\
  &=& \phi_0+\frac{1}{\sqrt{12 \pi G}} 
       \log \left| \frac{(t-t_0)-k/ \sqrt{12 \pi G \rho_0^2}}
           {(t-t_0)+k/ \sqrt{12 \pi G \rho_0^2}} \right| .
\label{e54}
\end{eqnarray}

\setcounter{equation}{0}
\section{Effect on the Gravitational Potential}
\indent

In this section, we calculate the effect of the 
cosmological constant and the scalar matter to the 
gravitational potential. For the special case of 
i) no scalar matter or ii) no cosmological term, the 
exact solutions are well-known.

\subsection{Exact solution for special cases}
\underline{i) No scalar matter case}
\indent

In this case, we take the standard metric in the form 
\begin{equation}
   ds^2=-h(r)^2 {dt}^2+f(r)^2 {dr}^2 
   +\left[ r^2 ({d \theta}^2 
   +\sin{\theta}^2 {d \varphi}^2) \right]~,  
\label{e55}
\end{equation}

\noindent
and the exact solution is given by~\cite{Kottler} 
\begin{eqnarray}
  && h(r)^2=1-r_0/r-\Lambda r^2/3, 
\label{e56}\\
  && f(r)^2
  =\frac{1}{1-r_0/r-\Lambda r^2/3 } .
\label{e57}
\end{eqnarray}

\vskip 3mm
\noindent
\underline{ii) No cosmological term case}
\indent

In this case, we take the isotropic metric in the form 
\begin{equation}
   ds^2=-h_1(r)^2 {dt}^2+f_1(r)^2 
    \left[{dr}^2 
    +r^2 ({d \theta}^2 +\sin{\theta}^2 {d \varphi}^2) \right]~,  
\label{e58}
\end{equation}

\noindent
and the exact solution is given by~\cite{Buchdahl}
\begin{eqnarray}
  \phi(r)&=&\phi_0 \log \left(\frac{r-r_0}{r+r_0}\right),
\label{e59}\\
  h_1(r)^2&=& \left(\frac{r-r_0}{r+r_0}\right)^{2 C},
\label{e60}\\
  f_1(r)^2&=& \left( 1-\frac{{r_0}^2}{r^2} \right)^{2}
             \left( \frac{r+r_0}{r-r_0}\right)^{2 C},
\label{e61}
\end{eqnarray}

\noindent
where
\begin{eqnarray}
 \phi_0&=&\sqrt{ \frac{2(1-C^2)}{8 \pi G}}, \quad C={\rm const.}\ .
      \nonumber
\end{eqnarray}

\vskip 3mm
   
\subsection{Cosmological term and scalar matter co-existing case}
\indent

When the cosmological term and scalar matter co-exist, we cannot 
solve analytically, and we calculate the effect on the gravitational 
potential approximately. For this purpose, we take the standard metric 
Eq.(\ref{e55}) and the equations of motion Eqs.({\ref{e2}) and 
(\ref{e3}) are given by 
\begin{eqnarray}
  \frac{4 r f^{'}}{f^2} +2 f-\frac{2}{f}-2 \Lambda r^2 f
     &=&\frac{ 8 \pi r^2 G {\phi^{'}}^2 }{f},
\label{e62} \\
  -\frac{4 r h^{'}}{f^2} -\frac{2 h}{f^2}+2 h-2 \Lambda r^2 h
     &=&-\frac{ 8 \pi r^2 G h {\phi^{'}}^2 }{f^2},
\label{e63} \\
   \left( \frac{ r^2 h\phi^{'}}{f} \right)^{'} &=&0.
\label{e64}
\end{eqnarray}

\noindent
These can be rewritten into the form 
\begin{eqnarray}
  \frac{f^{'}}{f}-\frac{h^{'}}{h}+\frac{f^2}{r}-\frac{1}{r}
   &=& \Lambda r f^2 ,
\label{e65} \\
  \frac{f^{'}}{f}+\frac{h^{'}}{h}&=&4 \pi G r {\phi^{'}}^2 ,
\label{e66} \\
  \left( \frac{ r^2 h\phi^{'}}{f} \right)^{'} &=&0.
\label{e67}
\end{eqnarray}

\noindent
From Eq.(\ref{e67}), we have 
$\phi^{'}= \alpha f/ r^2 h$ with constant $\alpha$.
Then we have
\begin{eqnarray}
  \left( \log{f/h} \right)^{'}=\frac{f^{'}}{f}-\frac{h^{'}}{h} 
     &=&-\frac{f^2}{r}+\frac{1}{r}+\Lambda r f^2,
\label{e68} \\
  \left( \log {f h} \right)^{'}=\frac{f^{'}}{f}+\frac{h^{'}}{h}
     &=&\frac{4 \pi G {\alpha}^2 f^2}{ r^3 h^2 }.
\label{e69}
\end{eqnarray}

We introduce the new variables $X$, $Y$ in the form 
$ \exp(X)=fh $, $ \exp(Y)=f/h $, then the above equation becomes 
in the form 
\begin{eqnarray}
  && Y^{'}=-\frac{\exp(X+Y)}{r}+\frac{1}{r}+\Lambda r \exp(X+Y),
\label{e70} \\
  && X^{'}=\frac{4 \pi G {\alpha}^2 \exp(2Y)}{r^3}.
\label{e71}
\end{eqnarray}

\noindent
For $\alpha=0$ case (no scalar matter case), we have the 
solution 
\begin{eqnarray}
   X=0, \quad \exp(Y)=\frac{1}{1-r_0/r-\Lambda r^2/3},
\label{e72}
\end{eqnarray}

\noindent
which is the exact solution to Eqs.(\ref{e56}) and (\ref{e57}).
Then we calculate the gravitational potential by considering
the region of $r$ where 
$r_0/r, \Lambda r^2, 4 \pi G {\alpha}^2/r^2 \ll 1$.
In this approximation, we have 
\begin{eqnarray}
  Y &\approx& \frac{r_0}{r}+\frac{\Lambda r^2}{3},
\label{e73}\\
  X &\approx& 2 \pi G {\alpha}^2 
  \left(\frac{1}{{r_1}^2}-\frac{1}{r^2}\right)
\label{e74}
\end{eqnarray}

\noindent
from Eqs.(\ref{e71}) and (\ref{e72}) where $r_1$ is constant.

In order to find the solution for $ \alpha \ne 0$, we put 
$r_0 {\rm (=const.)}  \rightarrow r_0(r)\ ({\rm function \ of \ } r )$.
Then Eq.(\ref{e70}) becomes 
\begin{eqnarray}
  Y^{'}&=&\left( \frac{r_0(r)}{r}+\frac{\Lambda r^2}{3} 
    \right)^{'}=-\frac{r_0}{r^2}+\frac{{r_0}^{'}}{r}
        +\frac{2 \Lambda  r}{3} \nonumber \\
  &\approx& -\frac{r_0}{r^2}+\frac{2 \Lambda r}{3}-\frac{X}{r}
\label{e75}
\end{eqnarray}

\noindent
which gives ${r_0}^{'}=-X$. Using Eq.(\ref{e74}), we have 
\begin{eqnarray}
 r_0(r)=r_2-2 \pi G {\alpha}^2 \left( \frac{r}{{r_1}^2}+\frac{1}{r}
       \right) ,
\label{e76}
\end{eqnarray}

\noindent
where $r_2$ is constant.

The gravitational potential $\Phi$ is given by
\begin{eqnarray}
 g_{00}&=&-(1+2 \Phi)=-\exp(X-Y) \nonumber\\
 &\approx&-(1-\frac{r_0(r)}{r}-\frac{\Lambda r^2}{3}+X) \nonumber\\
 &\approx&-(1+\frac{4 \pi G {\alpha}^2 }{r_1^2}
  -\frac{r_2}{r}-\frac{\Lambda r^2}{3}),
\label{e77}
\end{eqnarray}

\noindent
which gives 
$\Phi=2 \pi G {\alpha}^2/r_1^2-r_2/2r
-\Lambda r^2/6$. 
Therefore the scalar matter does not contribute to the 
gravitational force $F_r=-\partial{\Phi}/\partial r
=-r_2/2r^2+\Lambda r/3$
within our approximation. The cosmological term contribute to the 
repulsive force within the approximation. 

\setcounter{equation}{0}
\section{Summary and Discussion}
\indent

We consider the scalar field as the candidate of the dark matter. 
Then, in order to give the standard scenario of the astrophysics,
we study the Einstein theory with minimally coupled scalar field 
and the cosmological constant.
We have studied various classical solutions with minimally 
coupled scalar and the cosmological term in the cosmological, the 
galactic or solar scale. We obtained the exact solution
in the cosmology scale, where the scale factor expand in the power
law in the first beginning and then expand exponentially.
In the galactic or solar scale, we cannot find the exact solution, and 
examine the contribution from the scalar field to the gravitational 
potential and find that the scalar field does not contribute to the 
gravitational force within our approximation. In this way, 
in the cosmological scale, the scalar
field play the role of the dark matter in some sense. While, in the 
galactic or solar scale, the scalar field does not pay the role of 
the dark matter.

For the ordinary matter, we first start from the classical 
Lagrangian and quantize the field and treat it as the classical 
smeared matter and make the perfect fluid approximation. 
While, in our approach, we treat the scalar field as the classical 
field in the same level as the classical gravitational field. 
If the metric is homogeneous, it may give the same effect whether we 
treat the scalar field as the classical field or the quantized 
and classically smeared matter. But if the metric is not homogeneous and 
is space-dependent, there is the quite big gap in the step of the 
quantization and the treatment of the classically smeared matter.
In this sense, the scalar field may give the contribution to the 
gravitational force if we treat the scalar field as the quantized 
matter field.

\vskip 10mm

\noindent
{\large \bf Acknowledgement}:\\
Two of us (K.S. and K.U.) are grateful to the academic research funds
of Tezukayama University.

\newpage


\end{document}